# Reflections of Cultural Wealth:
# Exploring Identity in Physics through Photo Elicitation

Zosia Krusberg[1]
The University of Chicago

## Abstract

This paper presents a photo elicitation project that invites students to explore their identities in STEM and surface the cultural wealth they bring into the physics classroom. Grounded in critical race theory, educational theory, and affective neuroscience, the project asks students to take original photographs representing aspects of their lived experience—such as motivation, belonging, and support systems—and to write short reflections connecting these images to their journey in science. The activity affirms students' backgrounds and values as meaningful resources for learning physics, while also fostering connection, inclusion, and reflective engagement. Practical guidance for implementation is provided. This adaptable, low-barrier practice offers an accessible entry point for instructors seeking to humanize physics education in powerful and inclusive ways.

---

[1] Email: zosia@uchicago.edu

Introduction

In physics classrooms, students learn new conceptual frameworks, practice solving problems, and engage with their peers and instructors. However, they are rarely invited to reflect on who they are, what brought them to physics, and what sustains their engagement. As educators, we often overlook the deeply human dimensions of learning: students' motivations, identities, cultural histories, and personal journeys. This short classroom activity, adapted from a social work practice known as *photo elicitation*, invites students to explore and express their identities in STEM through image and reflection. In doing so, they surface the resources they bring into the classroom and also build a powerful foundation for voice, connection, and belonging.

Centering Cultural and Community Assets

One of the central goals of this project is to affirm the value of students' lived experiences—their families, languages, communities, histories, and motivations—as meaningful resources for learning physics. This perspective draws on asset-based educational frameworks such as *funds of knowledge* and *community cultural wealth* that challenge the assumption that students enter STEM classrooms with deficits to be remediated. Instead, these frameworks emphasize the rich knowledge, relationships, and strengths that students already possess. A key foundation for this perspective is *critical race theory*, a framework that invites educators to critically examine how structural inequality shapes student experience.

Critical race theory (CRT) originated in legal studies and was later adapted to education by scholars like Gloria Ladson-Billings and William Tate (1995). CRT recognizes that racism is not just a matter of individual bias but is also embedded in laws, institutions, and everyday practices—even those that appear neutral or colorblind. In the context of education, CRT highlights how opportunity and access are shaped by race and power, and affirms the importance of including the voices and stories of students who have been marginalized by dominant narratives. CRT encourages us to examine how systems uphold racial inequities, while also naming the agency, resistance, and cultural wealth that students of color bring to educational spaces (Delgado & Stefancic, 2017).

Building on the funds of knowledge framework developed by Moll and colleagues (1992), which emphasizes the everyday knowledge and practices students bring from their households and communities, Tara Yosso's (2005) community cultural wealth framework offers a more explicitly race-conscious model rooted in CRT. Yosso identifies six forms of capital that students from marginalized backgrounds often draw on to navigate and succeed in educational environments: aspirational, linguistic, familial, social, navigational, and resistant capital. These forms of capital often go unrecognized in physics classrooms, where technical proficiency and objectivity are emphasized over personal context. Yet when students submit a photo of a parent's tools, a quiet park bench where they study, or a religious symbol that grounds them, they are bringing these powerful forms of cultural wealth into view. This project offers a space where such assets are not just acknowledged but celebrated.

## Photo Elicitation as Pedagogy

Photo elicitation is a research practice in which photographs are incorporated into interviews or reflection processes to elicit deeper insight and emotional engagement (Harper, 2002). Used in clinical and qualitative research, it draws on the power of visual imagery to access memory and meaning that may not surface through language alone. From a neuroscientific perspective, visual processing engages evolutionarily older brain regions—including the occipital lobe and limbic system—that are more directly connected to emotion and memory than the prefrontal cortex, which governs verbal reasoning. This neurological distinction helps explain why photographs can evoke powerful emotional responses and open doors to deeper self-reflection, especially when they are used to surface experiences not yet fully formed in language (Gee, 2000; Immordino-Yang & Damasio, 2007).

In a physics classroom, photo elicitation offers an unusual but resonant pedagogical tool. It invites students to connect personal experience with disciplinary identity, and, in doing so, bridges a gap between physics as abstract and physics as human.

## The Classroom Activity

In this project, students are invited to create a visual and written reflection on their identity in STEM. They are asked to capture **five original photographs** that represent different facets of their experience, choosing from prompts such as:

- Your educational experiences in STEM
- Your sense of belonging (or exclusion) in STEM
- Your motivations for studying science
- Your role models or sources of inspiration
- Your support systems
- Your cultural background or identity
- Your experiences of challenge, failure, or resilience
- The advantages or disadvantages that shape your experience

For each image, students write a short reflection—approximately **200 words**—describing what is captured and how it connects to the selected theme or aspect of their identity in STEM. Students may also choose to explore other forms of expression—such as poetry, drawing, collage, or layered text and image—to more fully represent their experiences.

Importantly, students are encouraged to take **new** photos for the project rather than using existing images. This act of intentional noticing invites them to re-engage with their present environment and personal narrative.

## Expanding the Medium

Over time, many students have extended the project beyond the original photo-and-paragraph format. Some have included poetry, visual art, or layered collages that combine text and image to express more nuanced or abstract aspects of their identity. Others have photographed a drawing, created digital compositions, or presented diptychs that contrast challenges with sources of strength. These creative variations are welcome, and often the most powerful. They remind us that reflective engagement is not one-size-fits-all, and that identity, like science, can be expressed in many languages.

Reflections and Impact

Students have responded to this project with thoughtfulness, vulnerability, and authenticity. Many describe it as one of the most personally meaningful assignments of the course. The submissions vary widely in tone and form—some are documentary and literal, others symbolic, poetic, or metaphorical—yet all reveal something essential about how students see themselves in relation to STEM.

Some photos depict everyday moments of community and motivation: friends studying at a local donut shop, or a stethoscope resting beside a stack of physics notes, evoking both purpose and care. Others highlight recognition and inclusion, such as a student's name written on a whiteboard in a research lab, or the image of a mentor who is openly queer in STEM, whose presence served as a source of validation and inspiration for a transgender student.

Students often drew on family history and cultural roots. One captured their father's sacrifice in emigrating from El Salvador, another described working with a medical charity in Vietnam founded by their grandfather, linking past and future. A third-grade art project portraying a parent as a superhero offered a glimpse into the earliest seeds of admiration and aspiration.

Many submissions were deeply symbolic. One student submitted an image of a Thai iced tea cheesecake—flavorful but cracked—to express that imperfection does not diminish accomplishment. Another shared a photo of a candle casting a long shadow, writing that they had long felt themselves in the shadows of more accomplished peers, yet were beginning to kindle their own light.

Some pieces documented challenge and growth. One student shared an image of themselves doing math homework despite having dyscalculia, writing that overcoming their fear of math became a turning point in self-confidence. Others reflected on "failures" during the transition to college, including exams that shook their sense of capability, but also taught them resilience and perspective.

Several students took creative approaches to the assignment: one submitted a fully hand-crafted portfolio, integrating printed images, handwritten text, and original illustrations that tied their themes together visually. Others paired photographs with poetry, using verse to give voice to emotions that might otherwise be difficult to articulate.

For many students, the project offered a moment to reconnect with purpose. Some have written to me years later to say it was a highlight of the course, reminding them both of why they chose this path as well as of the people and experiences that continue to sustain them.

For instructors, these reflections offer rare and salient insight into the inner lives of students—insight that remains invisible in most physics assessments. Personally, I feel honored to be invited into these glimpses of students' worlds: their courage, creativity, longing, and strength. This project has reminded me, again and again, of the complexity and richness each student brings into the room—and of the possibility that even a small assignment can open a door to deeper connection.

## Suggestions for Instructors

For instructors interested in implementing this project into their courses, here are a few considerations:

- **Timing**: Assign the activity midway through the term, once students feel safe and connected but before end-of-quarter stress.
- **Grading**: Assess for completeness and depth of engagement, rather than technical skill in photography or writing.
- **Sharing**: Respect privacy and permit students to opt out of sharing their work. Participation in any class discussion or presentation should be fully voluntary.
- **Format**: Encourage alternative formats—photos with poetry, drawings, or short audio recordings—especially for students drawn to creative expression.
- **Flexibility**: Allow students to include one previously taken photo if it holds personal meaning; this small adaptation can deepen reflection while maintaining the spirit of the practice.

This project can be used in any STEM course and is especially valuable for instructors who wish to bring more reflection or human connection into their teaching. It works well as a stand-alone activity and can serve as an accessible entry point into more humanistic or inclusive approaches to physics education.

## Conclusion

*Reflections of Cultural Wealth* is a small assignment with deep resonance. It opens a door for students to see themselves and one another as complex, resourceful, and resilient human beings. It invites us, as instructors, to listen with new ears and see with new eyes. In doing so, we affirm that physics is not just about what students know, but also about who they are, and who they are becoming.